\documentclass[conference]{IEEEtran}
\IEEEoverridecommandlockouts
\usepackage{cite}
\usepackage{amsmath,amssymb,amsfonts}
\usepackage{algorithmic}
\usepackage{graphicx}
\usepackage{pgfplots}
\pgfplotsset{compat=1.18}
\usepackage{subcaption}

\usepackage{textcomp}
\usepackage{xcolor}
\usepackage{url}

\def\BibTeX{{\rm B\kern-.05em{\sc i\kern-.025em b}\kern-.08em
    T\kern-.1667em\lower.7ex\hbox{E}\kern-.125emX}}

\usepackage{newtxtext}       %
\usepackage{newtxmath}       
\usepackage{amsmath}

\usepackage{listings}
\usepackage{multirow}
\usepackage{soul}
\usepackage{xspace}
\usepackage{tikz}
\newcommand*\circled[1]{\tikz[baseline=(char.base)]{
            \node[shape=circle,draw,inner sep=1pt] (char) {#1};}}
            
\definecolor{gr}{rgb}{0.0, 0.5, 0.0}
\definecolor{b}{rgb}{0.0, 0.0, 1.0}
\definecolor{lb}{rgb}{0.0, 0.0, 0.5}
\definecolor{purple}{rgb}{0.55, 0.0, 0.55}
\definecolor{eclipseBlue}{RGB}{42,0.0,255}
\definecolor{eclipseGreen}{RGB}{63,127,95}
\definecolor{eclipsePurple}{RGB}{127,0,85}
\definecolor{lbcolor}{rgb}{0.9,0.9,0.9}  
\definecolor{lightblue}{rgb}{0.68, 0.85, 0.9}
\definecolor{k}{rgb}{0.0, 0.5, 0.0}
\definecolor{v}{rgb}{0.0, 0.0, 0.5}
\definecolor{b}{rgb}{0.0, 0.0, 1.0}

\definecolor{delim}{RGB}{20,105,176}
\definecolor{numb}{RGB}{106, 109, 32}
\definecolor{string}{rgb}{0.64,0.08,0.08}

\lstdefinelanguage{json}{
    showstringspaces=false,
    morestring=[b]",
    stringstyle=\color{delim},
    literate=
      {\{}{{{\color{string}{\{}}}}{1}
      {\}}{{{\color{string}{\}}}}}{1}
      {[}{{{\color{string}{[}}}}{1}
      {]}{{{\color{string}{]}}}}{1},
}

\lstdefinelanguage{SPARQL}{
  showstringspaces=false,
  comment=[l]{//},
  morecomment=[s]{/*}{*/},
  commentstyle=\color{gr}\ttfamily,
  morecomment=[l]{\#},       
  morecomment=[n][\color{lb}]{<http}{>}, 
  morestring=[b][\color{green}]{\"},  
  classoffset=1,
  morekeywords={?O, sh, sosa, owl, xsd, purl, ssr, prov}, keywordstyle=\color{b},
  classoffset=2,
  morekeywords={
    SELECT,CONSTRUCT,DESCRIBE,ASK,WHERE,FROM,NAMED,PREFIX,BASE,OPTIONAL,
    FILTER,GRAPH,LIMIT,OFFSET,SERVICE,UNION,EXISTS,NOT,BINDINGS,MINUS,
    STREAM, NAF, WINDOW, RANGE, ON, as, GROUP, BY, HAVING, ORDER, COUNT
  },
  keywordstyle=\color{gr}\textbf,
  classoffset=3,
  morekeywords={
   det, det1, det2, det3, det4, det5, det8, det11, car, hasScore, resultTime, madeObservation, wasGeneratedBy,
   isSampleOf, usedProcedure, Kalman, isResultOf, cam, madeBySensor, Sampling, Observation, hasResult, detectedIn, hasSimpleResult, isDetectionOf,
   image2, Image2D, obs2, Yolo, KalmanFilter,
   b1,b2,b3,b4,b5,b6,b7,b8,b9,b10,b11,
   trk1, trk2, trk3, trk,trk23,trk5,
   rule_w_1,rule_w_2, rule_w_3, rule, leaves, inFOV, rule_w_4, vMatch, ends,
   NodeShape, FoV, CQELSRule, prefixes, construct, rule_w_5, score,
   enters, trklet
  },
  keywordstyle=\color{purple},
  classoffset=4,
  otherkeywords={},
  morekeywords={
    window, sec, iou, @, &&
  },
  keywordstyle=\color{red},
  classoffset=5,
  morekeywords={a}, keywordstyle=\color{lb}\textbf,
}

\lstdefinelanguage{asp}{
  keywords={not, at,CONSTRUCT,WHERE,NAF,STREAM,window,FILTER,from,in, sosa},
  otherkeywords = {:-,@,[, ]},
  keywordstyle={\color{blue}},
  ndkeywords={pred, leaves, det, iSO, trajectory, enters,trklet, ends, starts, reid, inFoV, trk, iou,  score, vMatch,hasConfScore,isSampleOf, initiates, terminates},
  ndkeywordstyle=\color{lb},
  sensitive=false,
  comment=[l]{//},
  morecomment=[s]{/*}{*/},
  commentstyle=\color{gr}\ttfamily,
  stringstyle=\color{red}\ttfamily,
  morestring=[b]',
  classoffset=2,
  keywordstyle=\color{red},
  morekeywords={car},
}
\newcommand{\acron}{Thoth}%
\newcommand{\system}{\acron SP}%
\newcommand{\swarm}{$\acron\!\divideontimes\!{swarm}$}
\newcommand{\runtime}{$\acron\!\triangleright\!runtime$}%
\newcommand{\node}{$\acron\!\cdot\!node$}%
\newcommand{\name}[1]{\mbox{\ttfamily{#1}}\xspace}%

\IEEEoverridecommandlockouts\IEEEpubid{\makebox[\columnwidth]{ 979-8-3503-0322-3/23/\$31.00 $\copyright$2023 IEEE \hfill}\hspace{\columnsep}\makebox[\columnwidth]{ }}
\begin{document}

\title{Semantic Programming for Device-Edge-Cloud Continuum}

\author{\IEEEauthorblockN{Anh Le-Tuan}
\IEEEauthorblockA{\textit{Technical University Berlin} \\
Berlin, Germany 
\\
}
\and
\IEEEauthorblockN{David Bowden}
\IEEEauthorblockA{\textit{Dell Technologies} \\
Cork, Ireland 
}
\and
\IEEEauthorblockN{Danh Le-Phuoc}
\IEEEauthorblockA{\textit{Technical University Berlin} \\
Berlin, Germany 
\\
}

}

\maketitle

\begin{abstract}
This position paper presents \system, a Semantic Programming framework with the aim of lowering the coding effort in building smart applications on the Device-Edge-Cloud continuum by leveraging semantic knowledge.  It introduces a novel neural-symbolic stream fusion mechanism, which enables the specification of data fusion pipelines via declarative rules, with degrees of learnable probabilistic weights. Moreover, it includes an adaptive federator that allows the \runtime~to be distributed across multiple compute nodes in a network, and to coordinate their resources to collaboratively process tasks by delegating partial workloads to their peers. To demonstrate \system's capability, we report a case study on a distributed camera network to show  \system's behaviour against a traditional edge-cloud setup. 

\end{abstract}

\begin{IEEEkeywords}
edge computing, autonomous system, distributed AI, semantic computing
\end{IEEEkeywords}

\section{Introduction and Motivation}

The rapid advancements in machine learning, particularly \textit{deep neural network} (DNN), have led to a significant increase in the volume of multi-modal stream data that can be processed with remarkable accuracy. However, the integration of closed-loop "Smart-X" systems, such as Smart Factories, Smart Junctions, and Advanced Driving Assistant Systems (ADAS) in the SmartEdge project
~\cite{smartedge}
cannot afford the latency associated with cloud-based data ingestion and processing. Additionally, the sheer magnitude of multi-modal streaming data, such as camera and lidar data, makes the conventional approach economically infeasible, even with advancements in GPU/CPU and storage technology.

\begin{figure}[ht!]
    \centering
    \includegraphics[width=0.4\textwidth]{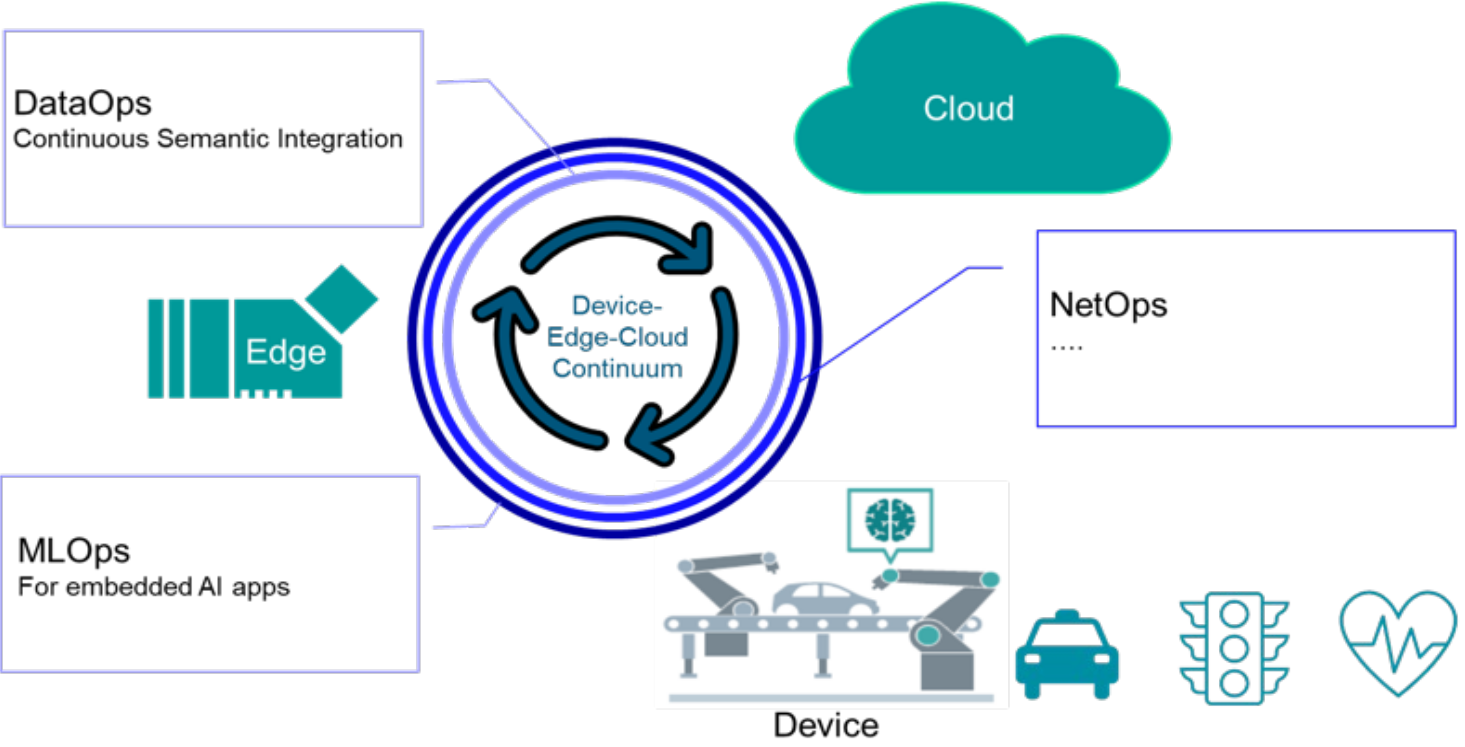}
    \vspace{-3mm}
    \caption{Device-Edge-Cloud Continuum.}
    \label{fig:continuum}
\end{figure}

This has motivated our endeavor to tackle the pressing challenge on how to design, program and implement a highly distributed network of device-edge-cloud continuum that takes part in dynamic information flows driven by unprecedented application logic. Our approach aims to facilitate the autonomous collaboration of edge devices operating in a dynamically changing environment. To accomplish this, significant enhancements are required to support the DataOps, NetOps and MLOps across the entire Device-Edge-Cloud continuum, illustrated in Figure~\ref{fig:continuum}. To enable this approach, we propose a Semantic Programming framework, \textbf{\system}, that leverages semantic knowledge to lower effort in programming DataOps, NetOps and MLOps agnostic to hardware and infrastructure. 

For DataOps, \system~ employ Semantic Streams~\cite{Le-Phuoc2020} powered by RDF stream processing (RSP) engines like CQELS engine\cite{Danh:2011} to solve the data integration problem across IoT systems. Semantic Streams facilitates the semantic interoperability among autonomous systems via shared formal semantics under standardized ontologies and vocabularies such as OGC/W3C Semantic Sensor Network Ontology (SSN)\cite{Armin:2019} and W3C WoT Thing Description (TD) Ontology~\cite{w3c2023wot}. 

Towards a seamless integration with DDN and multimodal stream data in MLOps, \system~ integrates a number of streaming data types, such as camera video streams and LiDAR point clouds, as well as supporting additional hardware platforms, such as ARM and mobiles~\cite{Anh:2018, Manh:2019}. Moreover,  \system~ also supports DNN-based data fusion operations~\cite{Danh:2021, Manh:2021}. Thus, \system~is designed to accommodate a DNN-based data stream fusion operations, which internally generates semantic stream that can be subsequently processed by the \runtime~in a unified data model, i.e. RDF-star\cite{rdfstar}. To support semantic interoperability in realtime, \runtime~ uses CQELS-QL, an extension of standadized query language SPARQL with advanced query patterns such event query patterns~\cite{Minh:2015, Daniele:2016} and probabilistic reasoning ~\cite{Danh:2021, Manh:2021} so that device, edge and cloud can coordinate a single query interface and language. Semantic streams also pave the way to unify NetOps with the semantic programming paradigm  following the vision of \textit{semantic communications}~\cite{DBLP:journals/corr/abs-2201-01389}.

\begin{figure*}
    \centering
    \includegraphics[width=.75\textwidth]{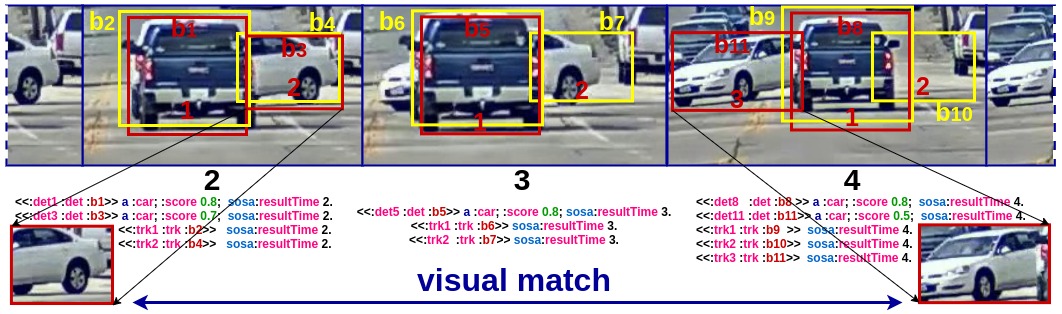}
   \caption{Semantic representation of multimodal data stream.}
    \label{fig:example}
\end{figure*}

\section{Semantic Programming In \system}
\label{sec:sempro}
This section introduces Semantic Programming (SP) paradigm for fusing multimodel data that draws inspiration from the human brain's semantic and episodic memory systems\cite{greenberg_verfaellie_2010}. The semantic memory refers to our brain’s repository of general world knowledge and episodic memory refers to our “episodic memory system”, which encodes, stores, and allows access to “episodic memories”, e.g. recollection of personally experienced events situated within a unique spatial and temporal contexts. By incorporating the principles of these cognitive processes into programming, SP aims to lower coding effort by exploiting semantic knowledge paired with human cognition. In this paradigm, programs are designed to manipulate and understand   information based on its \emph{semantic symbols} rather than relying solely on rigid algorithms or explicit instructions. 

To ground programming elements to semantic symbols understandable for both human and machine, we 
present sensory streams in RDF-star~\cite{rdfstar}. As illustrated in Figure~\ref{fig:example}, \system~fuses the video data stream observed by a camera (as a ${Sensor}$) as \emph{a stream of symbolic observations} representing these video frames. These frames are represented as instances of the class ${Image2D}$ that inherits from the generic ${Observation}$ class of SSN ontology~\cite{Armin:2019}. The detection of a video frame or a tracklet are represented as SSN $Sampling$, which are processed by a DNN model or a CV algorithm represented as a \emph{Procedure}.

\begin{figure}[ht!]
    \centering
    \includegraphics[width=.4\textwidth]{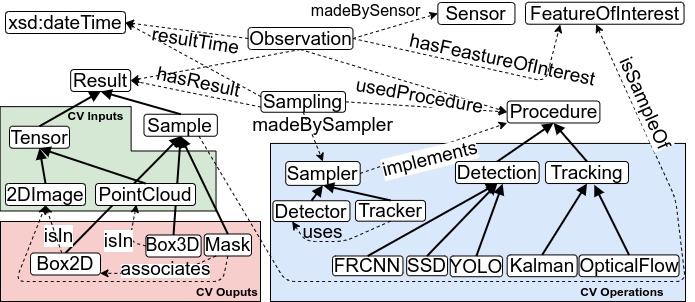}
    \caption{\small Data model of Semantic Streams based on SSN Ontology.}
    \label{fig:abstract}
\end{figure}

\begin{lstlisting}[
caption={\small{A Snapshot of Semantic Stream serialized with RDF-star.}}, 
label={rdf:frame2},
language=SPARQL,
numbers=left,
captionpos=b,
basicstyle=\scriptsize\ttfamily,  
backgroundcolor=\color{lbcolor},]
//time point/frame 2
<<:image2 a :Image2D>> a sosa:Observation; 
sosa:madeBySensor :cam1; sosa:resultTime 2.
//RDF-star descriptions of detection 1 and 3
<<:det1 :det :b1>> a :Detection; sosa:resultTime 2;
sosa:hasSimpleResult 'car'; :score '0.8';
:isDetectionOf :image2; sosa:usedProcedure :Yolo.
<<:det3 :det :b3>> a :Detection; sosa:resultTime 2;
sosa:hasSimpleResult 'car'; :score '0.7';
:isDetectionOf :image2; sosa:usedProcedure :Yolo.
//RDF-star descriptions of tracking 1 and 2
<<:trk1 :trk :b2>> a :Tracklet; sosa:resultTime 2;
sosa:usedProcedure :KalmanFilter.
<<:trk2 :trk :b4>> a :Tracklet; sosa:resultTime 2;
sosa:usedProcedure :KalmanFilter.
\end{lstlisting}

For example, Listing~\ref{rdf:frame2} presents a snapshot of semantic stream data in Figure~\ref{fig:example} where the red boxes are detected bounding boxes and the yellow boxes are tracked bounding boxes.
Line 5 denotes that the detection model generates an output consisting of a bounding box $b_1$, object type $car$, and confidence score $0.8$. Line 13 denotes that the bounding box $b_2$ is predicted by a Kalman filter and is tracked by tracklet 1.

Based on several theoretical and empirical studies of human perception considering \emph{"perception as hypotheses"},e.g\cite{Gregory:1980,Friston:2012}, a Semantic Stream Reasoning (SSR) program is formulated as a set of rules representing hypotheses from semantic streams above. Such rules can be written in 
CQELS-RL (an extension from CQEL-QL and SHACL) syntax. 
The formal semantics of SSR is specified in~\cite{Danh:2021}. In essence, there are two types of rules, \emph{hard rules} and \emph{soft rules}. The hard rule is used for background knowledge given by (non-monotonic) common-sense and domain knowledge that is regarded as "always true". The soft rules expresses association hypotheses with weights corresponding to probability degrees of these rules. 
Listing~\ref{soft-rule1} illustrates how a soft rule is written in CQELS-RL syntaxs based on the input stream illustrated in Listing~\ref{rdf:frame2}. This rule can be interpreted in plain English as \emph{"a vehicle enters a field of view if a bounding box of it has been detected first time in the last 5 seconds"}. 

\begin{lstlisting}[
caption={\small Soft rule to detect vehicles entering the Field of View.}, 
label={soft-rule1},
captionpos=b,
language=SPARQL,  
%numbers=left,
%xleftmargin=5.0ex
basicstyle=\scriptsize\ttfamily,  
backgroundcolor=\color{lbcolor},
escapeinside=||,
]
ssr:rule_w_1 a sh:NodeShape;
sh:rule [
  a sh:CQELSRule ;
  sh:prefixes ssr: ;
  sh:construct |"""|
  CONSTRUCT {<<?O :enters <ssr:FoV>> @ ?T.}
  WHERE { STREAM <:ssr> 
  { <<?Dt :det ?B >> @ ?T; :score ?S.
    ?B sosa:isSampleOf ?O ; a :car.
    FILTER (?S > 0.8)  }
    NAF STREAM <:ssr> window[5 sec] 
   {?O :inFOV ssr:FoV.} }
  |"""| ;
] ;
\end{lstlisting}

\begin{figure}[ht!]
    \centering
    \includegraphics[width=0.45\textwidth]{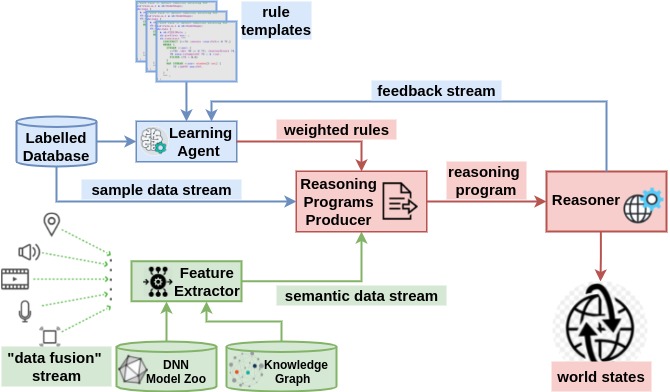}
    \caption{\small Learning and Fusion Workflow in \system.}
    \label{fig:concept}
\end{figure}

The weights of the soft rules are determined by the \name{Learning Agent} component in the starting phase the learning and fusion workflow illustrated in Figure~\ref{fig:concept}.
For the setup, the symbolic training samples are constructed from the labeled data and are fed to the \name{Learning Agent} component. The \name{Learning Agent} component computes a vector of weights for the soft rules, which are stored as rule templates, and passes them to the \name{Reasoning Programs Producer} component. For each training iteration, the \name{Reasoner} component returns a~\emph{feedback stream} of reasoning results back to \name{Learning Agent}. The \name{Learning Agent} then adjusts the weights of the soft rules until the answer sets, returned as a feedback streams, describe the most likely ground truth. 

The~\name{Feature Extractor} component detects the features of interest from the incoming data stream and maps them to their symbolic representation. The feature data is described with a~\emph{neuro-symbolic stream model}~\cite{Danh:2021} and its semantics are enriched via semantic knowledge graphs.

The~\name{Reasoning Programs Producer} component generates SSR programmes~\cite{Danh:2021}, which specify the fusion pipeline and the decision logic to choose the most likely state of the world at each evaluation. The reasoning program is evaluated by the \name{Reasoner} component, which employs an ASP solver. 

\section{Design of Semantic Programming Framework}
\subsection{Swarm Architecture}

The \system~ framework takes the input from a SSR program 
described in Section~\ref{sec:sempro}
. 
In essence, the resultant reasoning programs are an amalgamation of application logic and knowledge graphs, which specify the semantics of the inputs, outputs, and behavior of smart-devices in a \swarm. The SSR programs are then compiled and deployed to each smart-device's \runtime, called \node, which collectively form a \swarm~of devices collaborating to perform a common task. To allow flexibility in deploying a \swarm, they can be instantiated in one of two modes. 

If the devices in the swarm and their function are known at design time, the swarm can be predefined. This technique allows legacy devices, smart-devices, and clouds to be integrated into the same swarm. The characteristics of each device are stored in a common device repository in the form of a knowledge graph, which allows the function and connectivity of the devices to be statically bound into the reasoning program at design time. The reasoning programs are deployed to each smart-device's \runtime~during swarm formation. As legacy devices do not host the \runtime, they are integrated into the swarm by connecting to smart-devices. Sensors on the legacy devices are statically bound to smart-devices at design time, and their data streams processed on the smart-devices' \runtime. Clouds can be passive recipients of data or semantic streams, or actively participate in the swarm if they host a \runtime.     

If the \swarm~ must operate dynamically, i.e., with devices joining and leaving over its lifespan, the \node~ must act as autonomous agents that can join or leave the swarm on their own volition. All \node{s} host a \runtime, which runs a basic SSR program. 
The \swarm~ is instantiated with a single seed \node, which acts as the swarm coordinator (SC). The SC is running a special SSR program whose purpose is to form and manage the swarm's operation. The SC then enlists free smart-devices (SDs) into the swarm  to perform its task. When the a \node~ joins the swarm it is assigned a role in performing the task. To enable the \node~ to perform this role, the required SSR program is dynamically downloaded into the \system-runtime and executed. In other scenarios a free SD may request to join a swarm to make use of its services. The SDs use messages oriented middleware (MOM), such as DDS or V2X, and an customized network layer that allows them to interact with other devices in the swarm to collaborate on performing the task.         

The SC maintains the primary swarm state in its dynamic knowledge graph (DKG). It updates the swarm state when smart-devices join or leave the swarm, or when significant events occur. Other SDs in the swarm may maintain a partial copy of that DKG, depending on their role in the swarm and their computational and storage capacity. At a minimum a SD will know the characteristics and skills it possesses, as well as basic swarm state information, e.g., its role in the swarm.

Listing~\ref{ls:service-meta} illustrates a subscription message in JSON-LD format based on the WoT TD ontology~\cite{w3c2023wot}. Based on the semantic description provided by the subscribed nodes, the parent node can carry the stream discovery patterns which use a variable in the stream pattern, as shown in line 4 of Listing~\ref{subquery}. The variable \textit{?streamURI} then can be matched in other metadata. In this example, it is used to link with the sensors that generated this stream. Recursively, the subscription process can propagate the stream information upstream hierarchically, and vice versa, the discovery process can be recursively delegated to downstream nodes via sub-queries in CQELS-QL as a simple text message.

\begin{lstlisting}[
caption={\small Example of subscription message in JSON-LD.}, 
label={ls:service-meta},
language=json,
captionpos=b,
basicstyle=\scriptsize\ttfamily,  
backgroundcolor=\color{lbcolor},
escapeinside=||,
]
|"{\color{gr}\textbf{@context}}"|:"https://www.w3.org/2022/wot/td/v1.1",
|"{\color{purple}title}"|:"Camera2-At-Helsinki",
|"{\color{purple}id}"|:"urn:uuid:9489991a-7622-45b6-8437-f859835d4",
|"{\color{purple}description}"|:"Traffic Camera at Junction....",
|"{\color{purple}properties}"|: 
{|"{\color{purple}status}"|: 
 {|"{\color{purple}description}"|:"Stream Camera Feed at 30FPS",
  |"{\color{purple}type}"|: "string",
  |"{\color{purple}forms}"|:[{|"{\color{purple}op}"|:"readproperty",
          |"{\color{purple}href}"|:"RTSP://helsinki.fi/camera/2",
          |"{\color{purple}methodName}"|:"RTSP",
          |"{\color{purple}contentType}"|:"application/mp4" }],
  |"{\color{purple}readOnly}"|: |\color{b}{true}|}}
\end{lstlisting}

To form a swarm, it would require each \node~to operate as an autonomous agent which can collaborate with other peers to execute a distributed processing pipeline specified in CQELS-RL. An autonomous \node~can dynamically join a network of existing peers by subscribing itself to an existing node in the network, called a parent node, and it then notifies the parent node about the query service and streaming service it can provide to the network. These services can be semantically described by using vocabularies provided by VoCaLS~\cite{DBLP:conf/semweb/TommasiniSDBAPV18}. 
Hence, a subscription can be done by sending a RDF-based message via 
Websocket  or DDS in ROS2. 

To this end, when an autonomous \node~joins a network, it makes itself and its connected nodes discoverable and queryable to other nodes of the network. Moreover, each node can share its processing resources by executing a CQELS-QL query on it. This will help us treat a CQELS-QL query over \swarm~as a query to a sensor network. 
Note that network telemetry such as network failure, status and mirrored traffic streams, can be modeled as semantic streams to seamlessly integrate them into the control flow of a \swarm.

With the support of the above subscription and discovery operations, a SSR program written in CQELS-RL can be deployed across several sites,e.g., traffic cameras across a road network. Each \node~ gives access to data streams fed from streaming nodes connecting to it. Such stream nodes can ingest a range of sensors, such as air radar, loop detector and camera. When the stream data arrives, this node can partially process the data at its processing site, and then forward the results as tabular results or RDF stream elements to its parent node.

In this context, when a query is subscribed to the top-most node, called root node, it will divide this query to sub-query fragments and deploy at one or more sites via its subscribed nodes. A query fragment consists of one or more operators, and each fragment of the same query can be deployed on different processing nodes. Recursively, a sub-query delegated to a node can be federated to its subscribed nodes. All participant nodes of a processing pipeline can synchronise their processing timeline via a timing stream that is propagated from the root node. The execution process of sub-query fragments can use resources, i.e. CPU, memory, disk space and network bandwidth of participant nodes to process incoming RDF graphs or sets of solution mappings and generate output RDF graphs/sets of solution mappings. Output streams may be further processed by fragments of the same query, until results are sent to the query issuer at the root node. For example, the sub-query of the query in below Listing~\ref{subquery} can be sent down to the nodes closer to the streaming nodes, then the results will be recursively sent to upper nodes to carry out the partial COUNT query until it reaches the root node to carry out final computation steps to return the expected results.

\begin{lstlisting}[
caption={\small Example of Federated Query accross Cameras.}, 
label={subquery},
captionpos=b,
language=SPARQL,  
basicstyle=\scriptsize\ttfamily,  
backgroundcolor=\color{lbcolor},
escapeinside=||,
]
SELECT ?camera (COUNT(?truck) as ?truckCount)
WHERE{
 STREAM ?streamURI [RANGE 5m ON sosa:resultTime]
 {
   ?sensor a ssr:Camera; sosa:madeObservation ?obs.
   ?obs sosa:hasResult ?vFrame.
   ?truck a :Truck; ssr:detectedIn ?vFrame.
 }
 ?streamURI prov:wasGeneratedBy/a :TrafficCamera.
}
GROUP BY ?camera
HAVING (COUNT(?truck) > 1) 
ORDER BY ?truckCount
\end{lstlisting}

This federation process can be carried out dynamically thanks to the dynamic subscription and discovery capability above. Moreover, the processing topology  can be dynamically configured by changing where and how participant nodes subscribed themselves to the processing networks.The biggest advantage of this federation mechanism is the ability to dynamically push some processing operations closer to the streaming nodes to alleviate the network and processing bottlenecks which often happen at edge devices. Moreover, this mechanism can significantly improve the processing throughput by adding more processing nodes on demand.

\subsection{Component Design}

In this section, we present the design of components of the \system~ framework as illustrated in Figure~\ref{fig:concept}. 
\system~  provides a hardware-independent infrastructure to implement \system~ kernels for computing continuous
queries expressed in CQELS-QL. The \system-runtime accepts RDF streams as input and returns RDF streams or relational streams in the SPARQL format as output. \system~ allows for the creation of RDF streams by annotating extracted features from sensor input data streams. The output RDF streams can then be fed into any RSP engine, and the relational streams can be used by other relational stream processing systems.

\begin{figure}[ht!]
    \centering
    \includegraphics[width=.4\textwidth]{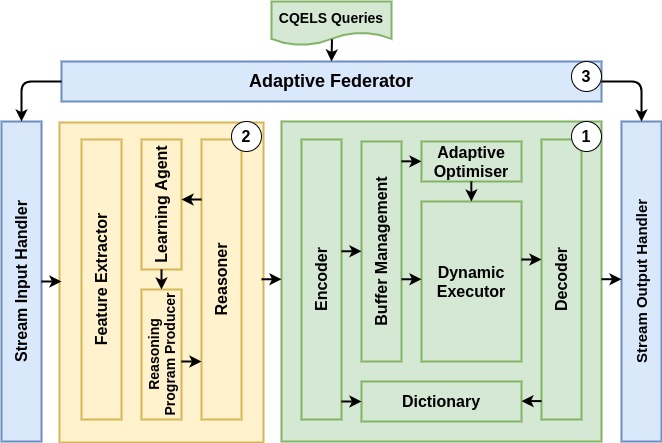}
    \vspace{-2mm}
    \caption{\small Component Design of \system.}
    \label{fig:cqels20}
\end{figure}

Figure~\ref{fig:cqels20} illustrates the component design of the \system~ framework. In general, \system~ consists of three subsystems.
Subsystem \circled{1}, the RSP processor,  extends the RSP primitives of its previous version~\cite{Danh:2011} to accelerate the grounding phase of a SSR. For example, multiway joins are used to accelerate the incremental ground techniques~\cite{Danh:2018}.
The second subsystem~\circled{2} is the SSR component as presented in Section~\ref{sec:sempro}. Finally, subsystem~\circled{3} is a stream adaptive federator, which is described in the remainder of the section.

To tailor the RDF-based data processing operations on edge devices (e.g, ARM CPU, Flash-storage), \system~ can be integrated with  RDF4Led~\cite{Anh:2018}, a RISC style RDF engine for lightweight edge devices, to build the Adaptive Federator based on~\cite{Manh:2019}. The each edge-based \runtime~ deployment is smaller than 10MB in size and needs only 4--6 MB of RAM to process millions of RDF triples on computationally constrained devices, such as BeagleBone~\cite{beagleboard},
Raspberry PI~\cite{pi}.
Therefore, \system~includes an adaptive federation mechanism to enable the coordination of different hardware resources, which in turn enables the construction of query processing pipelines by cooperatively delegating partial workloads to their peer agents. For elastically scale processing load the cloud infrastructure, \system~can use Apache Flink as the underlying software stacks for coordinating parallel execution processes using the approach in CQELS Cloud~\cite{Danh:2013}. 
This will lay the foundation for an integration of the adaptive optimizer with the cloud-based stream scheduler and operation allocations.

The Adaptive Federator acts as the query rewriter, which dynamically divides the input query into sub-queries. The rewriter then pushes the sub-queries as close to the streaming source device as possible by following the predicate push down approach used in common logical optimisation algorithms. The metadata subscribed to by the other \runtime~is stored locally. Similar to~\cite{Dell:2017}, such metadata allows the endpoint services of a \runtime~to be discovered via the Adaptive Federator. When the Adaptive Federator sends out a sub-query, it notifies the Stream Input Handler to subscribe and listens to the results returning from the sub-query. On the other hand, the Stream Output Handler sends out the sub-queries to other devices or sends back the results to the requester.


\section{A Case Study on Distributed Camera Network}
\subsection{Build a SSR program for Multi Object Tracking(MOT)}
In computer vision, a MOT pipeline is normally programmed in C/C++ or Python. 
With \system, the widely used tracking-by-detection approach~\cite{Ciaparrone:2019} can be emulated as an $\mathcal{M^{MOT}}$ in a declarative fashion with rules and queries, e.g. SORT in Listing~\ref{soft-rule2}. The key operations in tracking-by-detection approaches 
are as following: 1) detection of objects (using DDN-based detector), 2) propagating object states (location and velocity) into future frames, 3) associating current detection with existing.

Figure~\ref{fig:example} illustrates the SORT~\cite{Bewley:2016} algorithm which is a simple object tracking algorithm based on DNN detectors such as SSD~\cite{Liu:2016} or YOLO~\cite{Redmon:2017}. 
To associate resultant detections with existing targets, SORT uses a Kalman filter~\cite{Bishop:2001} to predict the new locations of targets in the current frame. 
At time point 2, the red boxes $b_1$ and $b_3$ are newly detected, and the yellow boxes $b_2$ and $b_4$ are predicted by a Kalman filter based on the tracked boxes from the previous frame. 
Then, the SORT algorithm computes an associative cost matrix between detections and targets based on the intersection-over-union (IOU) distance between each detection and all predicted bounding boxes from the existing tracklets.
In case some detection is associated with a target, the detected bounding box is used to update the target state via the Kalman filter. 
As in frame 2, the tracklets $trk_1$ and $trk_2$ are set to the two new bounding boxex $b_1$ and $b_2$ which are associated with predicted boxes $b_3$ and $b_4$ respectively. Otherwise, the target state is simply predicted without correction using the linear velocity model. For example, at time point 3, the detector misses detecting the white car due to an occlusion, however, the tracklet 2 is till assigned to box $b_7$ which contains part of the white car. 

To associate a detected bounding box $B$ with an object $O$, we use soft rules that assert the triple $\ll\!B~sosa\!:\!isSampleOf~O\!\gg$. 
Such rules can be used to represent {\em hypotheses}\/ on temporal relations among detected objects in video frames following a tracking trajectory. Particularly, the object's movement is consistent with the constant velocity model, e.g., the Kalman filter used in SORT~\cite{Bewley:2016}, and there is a detection associated with its trajectory, the fact $\ll\!B~sosa\!:\!isSampleOf~O\!\gg$ is generated. Here, $iou(B_1,B_2)$ states the IOU (intersection over union) condition of the bounding boxes $B_1$ and $B_2$ satisfy.

\begin{lstlisting}[
caption={\small A soft rule to emulate SORT algorithm~\cite{Bewley:2016}.}, 
label={soft-rule2},
captionpos=b,
language=SPARQL,  
numbers=left,
basicstyle=\scriptsize\ttfamily,  
backgroundcolor=\color{lbcolor},
escapeinside=||,
xleftmargin=5.0ex
]
ssr:rule_w_2 a sh:NodeShape ;
sh:rule [
  a sh:CQELSRule ;
  sh:prefixes ssr: ;
  sh:construct |"""|
  CONSTRUCT { ?B1 sosa:isSampleOf ?O. }
  WHERE { STREAM <:ssr> 
        { <<?Dt :det ?B2 >> @ ?T; :score ?S.
         <<?Trk :trk ?B1 >> @ ?T.
           ?Trk :trklet ?O. }
  FILTER (?S>0.8 && iou (?B1,?B2) > 0.8)}}|"""| ;
  ] ;
\end{lstlisting}

Similarly, we can also emulate DeepSORT~\cite{Wojke:2017} via a another soft rule
that can search for more supporting evidences to link a newly detected bounding boxes from an occluded tracklet using visual appearance associations, e.g. frames 2 and 4 of Figure~\ref{fig:example}. 
For this, one can search for pairs of bounding boxes from recently occluded tracklets w.r.t. visual appearance 
(more details in\cite{Danh:2021}).

\subsection{Federating a SSR Program in \swarm}

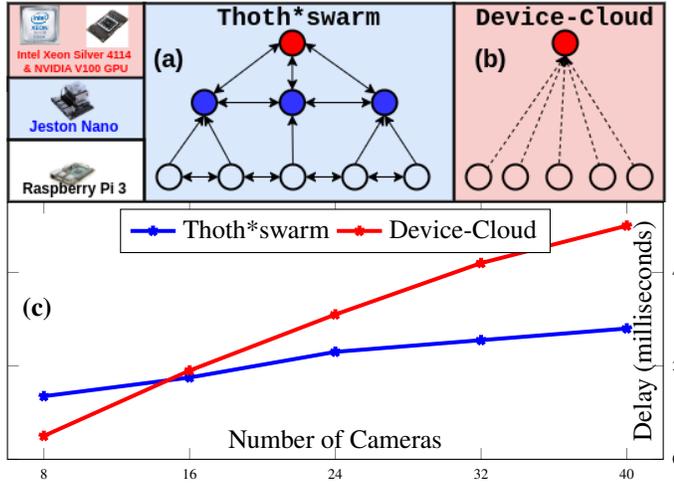
\begin{figure}[ht!]
\centering
\includegraphics[width=0.49\textwidth]{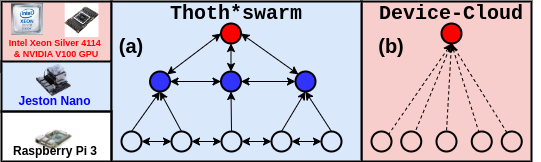}

\begin{tikzpicture}[scale=1]
\node[text width=1cm] at (.7, 2) {\textbf{(c)}};
\begin{axis}
    [   
        height=5cm,
        width=10.3cm,
        xlabel={Number of Cameras},
        x label style={at={(axis description cs:0.5,0.15)}},
        ylabel={Delay (milliseconds)},
        y label style={at={(axis description cs:0.94, 0.5)}},
        xmin=0, xmax=50, ymin=0, ymax=800,
        xtick = {0,    8,     16,     24,   32,   40,  48,  56},
        ytick = {0,  200,    400,   600},
        xmin = 6,
        xmax = 42,
        ymax = 550,
        xticklabel style = {font=\tiny},
        yticklabel style = {font=\tiny},
        yticklabel pos = {right},
        legend style={at={(0.5, 0.975)}, anchor=north,legend columns=-1},
    ]
\addplot[color=blue, mark=star, line width= 1.5pt]
coordinates {
(8,  135)
(16, 175)
(24, 230)
(32, 255)
(40, 280)
};
\addlegendentry{Thoth*swarm}

\addplot[color=red, mark=star, line width= 1.5pt]
coordinates {
(8,  50)
(16, 190)
(24, 310)
(32, 420)
(40, 500)
};
\addlegendentry{Device-Cloud}
\end{axis}
\end{tikzpicture}

\caption{\small Scaling behaviour of \swarm~vs. device-cloud approach.}
\label{fig:scaling}
\end{figure}

Next, we report the implementation of our case study on a distributed camera network. This network is built based on the data provided by the AI City Challenge (AIC)~\cite{DBLP:conf/cvpr/TangN0YBWKAH19}. The AIC has 40 cameras spanning across 10 intersections in a mid-sized US city. The baseline to compare the efficiency and scalability of \swarm~ is the traditional device-cloud (DC) infrastructure that ingests data from these 40 cameras connected by 40 roadside units (RSU) into a centralized server (the red node as shown in Figure~\ref{fig:scaling}(b)). The red node is a powerful server (2 Intel Xeon Silver 4114 Processors, 1TB RAM, V100 GPU cards with 16GB) representing for a cloud infrastructure. In this setup, the red node will handle all workload of the above $\mathcal{M^{MOT}}$ program including its SORT and DeepSORT algorithms.  An RSU runs on a Raspberry Pi 3 Model B which is used only for encoding/decoding video streams from its attached camera in this baseline. To simulate camera streams, we used these RSUs to replay the recorded data from AIC at a speed of 10 frames per second (fps) for each camera. 

Then, \swarm~setup adds 8 Jetson Nanos (JNs)~\cite{jetson}
to DC setup for offloading heavy processing load of DNN-based object detections  the DC setup, called device-edge-cloud (DEC). We use the Yolov5 pre-trained model for both DC and DEC setups. 
A group of RSUs will be connected to a JN via a wired network if they are geographically collocated to an intersection. 

When a $\mathcal{M^{MOT}}$ is registered to the red node of the \swarm, it delegates some subtasks to the blue nodes reprenting for Jetson Nano devices. For examples, subtasks on object detections and tracking are specified as in line 9 and line 10 of Listing~\ref{soft-rule2}.
As we can see in Figure~\ref{fig:scaling}(c), when the number of RSUs increases to more than 10, \swarm~starts to perform better than a DC counterpart in terms of latency. Note that, even DEC has more processing power than DC, but adding one more communication hop to the network topology will add more delay if the red node is not overloaded in terms processing or bandwidth. However, for the most heavy operation Yolov5, a V100 can process 100-250 fps, while each Jetson nano can process 10-25 fps. Consequently, the benefit of having edge node is getting clearer when having more processing load and network demand, i.e. streaming from more than 10 cameras.

\vspace{-2mm}

\section{Conclusion and Outlook}

The paper introduces the \system~ framework, aiming to materialize the SP paradigm, wherein developers can concentrate solely on the semantic symbols abstracting away the complexities of DataOps, NetOps, and MLOps on multimodal data streams. Hence, \system~ simplifies the development process with declarative queries and rules composed from such symbols. An initial implementation was presented, accompanied by a case study on a distributed camera network showing that a \swarm~ can dynamically offload the processing load to the edge devices closer to the sensing devices.

While the initial implementation of \system~ within the context of the Horizon Europe project, SmartEdge, has shown a promising scaling behavior and improved programability, there are several exciting and challenging endeavors we are eager to explore to ensure \system~works robustly in SmartEdge use cases, such as Smart Factories, Smart Traffic, and ADAS. The first challenge is effectively managing the dynamicity of the \swarm ~when a \node~ joins or leaves the peer-to-peer network (e.g., V2X or mobile robots) without disrupting any services that do not rely on it. Addressing this issue is crucial to maintain smooth operations in dynamic environments.

The second challenge involves developing optimization algorithms to adaptively optimize the federated computing pipeline on distributed streams of multi-modal data. This is essential to ensure optimal performance and resource utilization in the face of constantly changing data patterns and processing demands. Finally, we aim to automate the process of building SSR  programs using both semantic knowledge and large language models amid the current rapid development of AI technologies. Automating this process will expedite program development and enhance the efficiency of \system~in handling complex and diverse data and deployment scenarios.

\vspace{-2mm}
\section*{Acknowledgment}
\vspace{-1mm}
This work is supported by the German Research Foundation (DFG) under the COSMO project (grant No. 453130567), and by the European Union's Horizon RIA 
under the grant agreement No. 101092908 (SmartEdge) and  WINDERA programme under the grant agreement No. 101079214 (AIoTwin). 

\vspace{-2mm}
\bibliographystyle{IEEEtran}
\bibliography{main}

\end{document}